\documentclass[prl,twocolumn,showpacs,aps,superscriptaddress]{revtex4-1}
 
\usepackage{amsmath}
\usepackage{amssymb}
\usepackage{amsmath}
\usepackage{comment}
\usepackage{times}
\usepackage{color} 
\usepackage{xspace}
\usepackage{amssymb,amsmath}
\usepackage{braket}
\usepackage{tabularx}
\usepackage{mathtools}          
\usepackage{setspace}

\usepackage[unicode]{hyperref}
\hypersetup{
	pdftitle={On the theory of spin-1 BECs},
	pdfauthor={Jiangnan},
	pdfsubject={Spin1-theory},
	colorlinks=true,
	linkcolor=blue,
	citecolor=blue,
	filecolor=black,
	urlcolor=blue
}
\def\urlprefix{}
\def\url#1{}

\usepackage{bm}
\usepackage{color}

\newcommand{\be}{\begin{equation}}

\newcommand{\ee}{\end{equation}}

\newcommand{\bea}{\begin{eqnarray}}

\newcommand{\eea}{\end{eqnarray}}

\begin{document}

\title{Unified Geometric Perspective for Spin-1 Systems: Bridging Nematic Director and Majorana Stars}

\author{Jiangnan Biguo}
\email{biguojiangnan21@gscaep.ac.cn}
\affiliation{Graduate School of China Academy of Engineering Physics, Beijing 100193, China}
\author{Rourou Ma}
\affiliation{Interdisciplinary Center for Theoretical Study, University of Science and Technology of China, Hefei, Anhui 230026, China}
\begin{abstract}
We present a unified geometric approach for spin-1 systems that connects seemingly distinct geometric representations such as the nematic director, the Cartesian representation and the Majorana stellar representation. Starting from a product state of two distinguishable spin-1/2 particles, we provide a direct way to capture crucial geometric information. This perspective reveals the fundamental interplay between subspace projection and geometric constraints.  This approach effectively maps magnetic solitons onto a kink model, allowing us to derive their equations of motion, a task not readily achieved with traditional methods. This simplified dynamical description reveals that the novel transition of these solitons in a harmonic trap corresponds to a fundamental transformation between kink and dip structures in the underlying geometry.

\end{abstract}

\maketitle
\textit{Introduction.--}The Bloch sphere provides a complete and elegant geometric language for spin-1/2 systems \cite{Sakurai2020modern}. While a spin-1 state possesses four intrinsic degrees of freedom, the three standard spin operators provide only a partial description. To capture the full geometry, two primary paradigms have been developed, yet a significant gap remains between their conceptual intuition and operational utility. The first paradigm employs the quadrupole tensor $\mathcal{N}_{ij}$ \cite{Nematic_original, Ueda_hydrodynamic}. While it provides a rigorous algebraic foundation, it introduces a notable redundancy: its five independent components exceed the minimum set of parameters required, and its physical eigenvectors, the nematic directors $\vec{u}$ and $\vec{v}$, are embedded within the tensor, making their direct dynamical tracking difficult\cite{Nematic_original,DalibardNematic,BlakieVortexTwocomponent,Orientiational_order,Nematic_oribit_couple,Ueda_hydrodynamic}. The second paradigm, the Majorana stellar representation, elegantly maps the spin-1 state to two points on the Bloch sphere \cite{MajoranaPaper,KawaguchiSpin1,DiscreteSymmetry,DiscreteSymmetry2}. Although this representation offers an unparalleled geometric intuition, it remains an operational challenge: extracting the stars' coordinates  requires solving the roots of a characteristic polynomial, a process that lacks a direct correspondence with Hermitian operators. To bridge this gap, there is a need for a reconciling framework founded on three-dimensional vectorial observables, supported by direct Hermitian operators that can faithfully and effectively characterize the complete intrinsic information of a spin-1 state, enabling a reciprocal understanding where geometric intuition and algebraic rigor inform one another. 

In this Letter, we resolve this challenge by introducing a unified geometric framework founded on a simple but powerful concept. We begin by constructing the spin-1 state from a tensor product of two distinguishable spin-1/2 particles. The construction identifies the location of two Majorana stars to spin orientations of the two spin-1/2 particles. This perspective naturally decomposes the geometry of the two Majorana stars into an average vector $\vec{\beta}$ and a relative vector $\vec{\gamma}$, and  bridges the two existing paradigms: $\vec{\beta}$ is proportional to the spin vector $\mathbf{F}$, while $\vec{\gamma}$ is precisely the nematic director $\vec{u}$. Crucially, our approach enables the construction of a vector quantity, $\vec{\Gamma}$, which serves as a direct measure of the elusive relative vector $\vec{\gamma}$ and its dynamics. This product-state construction offers a generalizable paradigm for high-spin systems.

Furthermore, this framework clarifies the geometric constraints underlying the mapping of spin-1 systems onto effective two-component models. While specific spin-1 states can be mapped onto pseudo-spin-1/2 systems via spin rotations\cite{ChaiXiaoSO(3),XQYu_static_FDS,BlakieVortexTwocomponent}, a systematic geometric understanding of this correspondence has remained elusive. We demonstrate that this mapping is associated with the locking of the average vector's orientation.

The power of this formalism is demonstrated by applying it to magnetic solitons in antiferromagnetic condensates. We reveal that these solitons are fundamentally hidden kinks in the relative vector $\vec{\gamma}$, a structure invisible to standard density or spin probes. By deriving the equations of motion for $\vec{\Gamma}$, we provide the first direct description of this kink's dynamics. This reveals the underlying picture of a previously observed transition of these solitons in a harmonic trap \cite{RelativePhaseDomainwall}, reinterpreting it as a fundamental geometric transformation of the soliton's core between a dip and a kink.

\textit{Majorana representation and product state construction.--} The Majorana representation provides an excellent framework to capture the geometric nature of high-spin states. The core insight is that a spin-$F$ state can be viewed as the symmetric state of $2F$ spin-1/2 particles. For a spin-$F$ state $|\psi \rangle =\sum_{m_F =-F}^{F} \psi_{m_F} |F,m_F\rangle$,
 the Schwinger boson formalism employs commuting boson operators ($\hat{a}^\dagger, \hat{b}^\dagger$) to express $|\psi \rangle$ as\cite{Schwinger2001}
\begin{align}
    |\psi \rangle \propto \prod_{i=1}^{2F} \left(\cos \frac{\theta_i}{2}\hat{a}^\dagger + \sin \frac{\theta_i}{2}e^{i\phi_i}\hat{b}^\dagger \right)|0\rangle,
\end{align}
where $|0\rangle$ is the bosonic vacuum. The state is geometrically represented by the position vectors $\vec{p}_i = (\sin\theta_i \cos\phi_i, \sin\theta_i \sin\phi_i,\cos\theta_i),(i=1,2..,2F)$ of the stars. Here $\theta_i\in[0,\pi]$ is the polar angle and $\phi_i \in [0,2\pi)$ is the azimuthal angle of the $i$-th star in Bloch sphere. 

While the Majorana representation treats the $2F$ spin-1/2 particles as indistinguishable bosons, it  blocks the extraction of non-symmetric information. To overcome this drawback, we consider an insightful hypothetical construction from $2F$ distinguishable spin-1/2 particles first. For a given spin-$F$ state, we can conceptually construct $(2F)!$ different tensor product states from 2$F$ individual spin-1/2 states. Each of these 2$F$ spin-1/2 states maps to a unique point on the Bloch sphere. The collection of all such points forms a set identical to the Majorana stars, the only distinction being the inherent indistinguishability in the Majorana representation. Let's pick one such product state for discussion,
\begin{align}
    |\Phi \rangle = \bigotimes_{i=1}^{2F}(\cos\frac{\theta_i}{2}|\uparrow \rangle_i + \sin\frac{\theta_i}{2}e^{i\phi_i}|\downarrow \rangle_i).
\end{align}
Here $|\Phi\rangle$ is normalized while the norm of the original spin-$F$ part $|\psi\rangle$ is given by Wick's theorem  $\langle \psi | \psi\rangle = \sum_{\sigma}\prod_{i=1}^{2F} \langle \psi_i | \psi_{\sigma_i}\rangle$ \cite{GeometricalHydrodynamics}, where $|\psi_i\rangle = (\cos \frac{\theta_i}{2}|\uparrow \rangle_i + \sin\frac{\theta_i}{2}e^{i\phi_i}|\downarrow \rangle_i)$, and $\sigma$ is a permutation of $2F$ spin-1/2 states.  Operators within the spin-$F$ subspace correspond uniquely to the exchange symmetric operators, since the resulting state must remain within the spin-$F$ subspace (which is the maximally symmetric subspace of the 2$F$ spin-1/2 system). Specifically, such an operator $\hat{O}$ can be represented by a symmetrized product of individual Pauli matrices acting on each spin-1/2 particle:
\begin{align}
    \hat{O} = \mathcal{S} (\sigma_{n_1}^{(1)}\otimes \sigma_{n2}^{(2)}\otimes...\otimes \sigma_{n_{2F}}^{(2F)}),
\end{align}
where $\sigma_{n_k}^{(k)}$ is a Pauli matrix (or identity) acting on the $k$-th spin-1/2 particle, and $\mathcal{S}$ denotes the symmetrization operator over all particle indices.
The expectation value of such an operator on our chosen product state has a direct geometric interpretation: $\langle \hat{O}\rangle = \mathcal{S}(p_{n_1}^1 p_{n_2}^2 ... p_{n_{2F}}^{2F}),$
where $p_{n_k}^k$ refers to $n_k$-th component of the position vector $\vec{p}_k$ associated with $k$-th spin-1/2 particle, and $p_0^k = 1$(representing the identity operator for that spin).  Crucially, this expanded Hilbert space allows for the definition of non-symmetric operators, which represents the relative information between the Majorana stars, thus enabling us to characterize information beyond what is contained by operators solely within the spin-$F$ subspace. In the subsequent discussion, we will leverage this product state perspective to investigate specific problems within spin-1 theory.

\textit{Spin-1 theory.--} A spinor $\psi = (\psi_{+1}, \psi_0, \psi_{-1})^T$ describing a spin-1 state possesses four independent intrinsic degrees of freedom  beyond the global phase $\phi$ and total density $n \equiv \sum_m |\psi_m|^2$. This four-dimensionality necessitates descriptions beyond the three parameters provided by the expectation values of spin operators $\hat{S}_i$ (magnetic dipole moments). The standard method addresses this by employing the quadrupolar tensor density $\mathcal{N}_{ij}= \langle \hat{S}_i \hat{S}_j + \hat{S}_j \hat{S}_i\rangle /2 \quad i,j = x,y,z$ \cite{Sakurai2020modern}. Its eigenvectors $\{ \vec{u},\vec{v},\mathbf{F}\}$ define the principal axes of spin fluctuations and collectively describe the complete geometric information of the spinor $\psi$. While the magnetic spin vector $\mathbf{F}$ is given by the expectation values of spin operators $\langle \mathbf{\hat{S}}\rangle$. The method for capturing the nematic directors $\vec{u}$ and $\vec{v}$ utilizes a Cartesian representation\cite{Ohmi1998}, defined as:
\begin{eqnarray}
    \psi_x = \frac{\psi_{-1}-\psi_{+1}}{\sqrt{2}},\psi_y = \frac{-i(\psi_{+1}+\psi_{-1})}{\sqrt{2}},\psi_z = \psi_0.
    \label{Cartesian_rep}
\end{eqnarray}
In this representation, $\vec{\psi} = (\psi_x,\psi_y,\psi_z)^T$ behaves like a vector under SU(2) rotation. It's known that after extracting a proper phase rotation, we find $\vec{\psi} = e^{i\theta}(\vec{u}+i\vec{v})$. This relation is also used to characterize the phase ordering in antiferromagnetic systems\cite{BlakieNematic}.

Here we can introduce a physically motivated decomposition of the two unit position vectors $\vec{p}_1, \vec{p}_2$ introduced by Majorana representation: (a). The average vector $\vec{\beta} \equiv (\vec{p}_1 + \vec{p}_2)/2$ directly yields the magnetization direction: $\mathbf{F} \propto \vec{\beta}$\cite{BrunoNematic,GeometricalHydrodynamics}. (b). Crucially, the relative vector $\vec{\gamma}\equiv (\vec{p}_1 - \vec{p}_2)/2$ is orthogonal to $\vec{\beta}$ and encodes the fundamental orientation linking the Majorana stars to the quadrupolar tensor. Its direction identifies the principal axis $\vec{u}$ of $\mathcal{N}_{ij}$\cite{footnote1}(This relationship is also mentioned in \cite{RelativeVectorRelation}). By the Pythagorean theorem, their magnitudes are related by $|\vec{\beta}|^2 + |\vec{\gamma}|^2 =1$.

 The vector $\vec{\gamma}$ cannot be expressed directly as the expectation value of any operator solely within the spin-1 space, because exchanging the Majorana stars reverses $\vec{\gamma}$ while the wavefunction remains invariant. To capture this crucial relative information, we decompose the product state of two spin-1/2 particles into total spin basis 
 \begin{align}
     |\Phi \rangle = \psi_{+1} |1,+1\rangle + \psi_0 |1,0\rangle + \psi_{-1}|1,-1\rangle + \phi_0 |0,0\rangle.
 \end{align}
  Here, our original spin-1 state $|\psi \rangle$ corresponds to the triplet sector. We denote the magnitude of spin-1 state $n \equiv \langle \psi | \psi \rangle$. The singlet component $\phi_0$ is given by
  \begin{align}
    \phi_0 = \pm \sqrt{\psi_0^2 - 2\psi_{+1}\psi_{-1}},
  \end{align}
has a magnitude $\alpha \equiv |\phi_0|^2$, satisfying $\alpha^2 + |\mathbf{F}|^2 = n^2$\cite{footnote2}. The sign ambiguity fundamentally reflects the exchange anti-symmetry of the singlet state. The product state norm is given by $n_{\Phi} = n + \alpha$.By projecting the expectation values of the two-spin operators onto the spin-1 subspace, we find the total spin $\mathbf{F}$ is rescaled by this norm, giving $\mathbf{F} = n_\Phi \vec{\beta}$.  Crucially, we define a relative vector operator $\hat{\mathbf{\Gamma}} \equiv \frac{1}{2}(\hat{\bm{\sigma}} \otimes I - I \otimes \hat{\bm{\sigma}})$ acting on the two-spin space. Its expectation values provide the components of $\vec{\Gamma} = n_\Phi \vec{\gamma}$: 
  \begin{align}
  \nonumber
  \Gamma_x &= -(\psi_+^* \phi_0 -\psi_{-}^* \phi_0 + c.c. )/\sqrt{2},\\
\nonumber
  \Gamma_y &= i(\psi_+^* \phi_0 + \psi_{-}^* \phi_0 - c.c.)/\sqrt{2},\\
  \Gamma_z &= \psi_0^* \phi_0 + \psi_0 \phi_0^*,
  \label{gamma_define}
  \end{align}
where "c.c." stands for complex conjugate. The singlet density $|\phi_0|^2$ governs the magnitude of $\vec{\Gamma}$.

An illuminating form making use of Cartesian representation defined in Eq.~\eqref{Cartesian_rep} is $\vec{\Gamma} = \phi_0^* \vec{\psi} + \phi_0 \vec{\psi}^*$. Its counterpart, the vector $\vec{\kappa} = -i(\phi_0^* \vec{\psi} - \phi_0 \vec{\psi}^*)$, identifies the direction of $\vec{v}$. This ultimately leads to the Cartesian representation $\vec{\psi} = (\vec{\Gamma}+i\vec{\kappa})/2\phi_0^*$. For a general spin-1 state evolving along a closed loop $C$, the total geometric phase $\Phi_g$ decomposes into a sum of individual solid angles swept by the Majorana stars and a correlation term between two stars\cite{Berryphase4}: $\Phi_g = -\frac{1}{2} \sum_{i=1}^2 \Omega_i + \Phi_{\text{corr}}$, where $\Omega_i$ are standard single-particle solid angles contribution on Bloch sphere, the correlation term within our framework, takes the form 
\bea
\Phi_{\text{corr}}= -\oint \frac{n_\Phi \left( \vec{\beta}\times \vec{\gamma}\right) \cdot d \vec{\gamma}}{4n}.
\eea

The geometric framework also provides a clear interpretation for the widely used projection of spin-1 condensates onto effective two component models. For a spin-1 state, there are mainly two types of subspace projection. The first type is $\psi_{+1}=0(\psi_{-1}=0)$. The geometric interpretation is one of the Majorana stars is fixed at the south pole(north pole) while the other remains free to move. We specifically focus on the other case $\psi_0 = 0$. Expressing $\psi_0$ in terms of the Majorana star angles $(\theta_i,\phi_i)$ and imposing this condition reveals a geometric constraint: $\phi_1 = \phi_2 +\pi, \theta_1 = \theta_2$, meaning they are antipodally symmetric about $z$-axis on the Bloch sphere. Any spatially uniform spin-1 condensate can be projected onto a pseudo-spin-1/2 system by an appropriate SO(3) spin rotation. For spatially varying wavefunction $\psi_m(\textbf{r})$, if it has finite magnetization $(\mathbf{F}(\mathbf{r})\neq 0)$, the magnetization axis must be spatially collinear, ensuring that a global rotation can align it with the $z$-axis. Examples satisfying this condition include ferro-dark solitons\cite{XiaoquanYuFerrodark,footnote1}, rogue waves\cite{Yanzhenya} , half-quantum vortices\cite{BlakieNematic,BlakieVortexTwocomponent} and other solitary excitations\cite{BDBspin1Soliton,BrightDarkNospintwist,SpinToTwocomponent,PRA_dark_antidark_fds}. Without external magnetic field, the Hamiltonian's SO(3) symmetry ensures rotated solutions remain valid under spin-rotation\cite{ChaiXiaoSO(3),ZhangwenxianSpin1Soliton}. In the presence of external magnetic fields, spin rotation must be systematically treated as a spin-basis transformation\cite{Feynman_Lectures_3}. Here, the Hamiltonian itself (including field terms) must be unitarily transformed to yield an equivalent two-component description.

Another example is Bogoliubov spectrum relationship between spin-1 and pseudo-spin-1/2 condensates. There are three branches in spin-1 condensates\cite{BdGspin1}, since the correspondence of ground state, two of these branches are identical to two-component with coherent coupling\cite{TwoComponentBdG}. The other branch is spin-twist branch which exactly breaks the projection condition. The instability of spin-twist branch will lead to spontaneous domain formation\cite{Youlispin1BdG}. This connection is also reported in \cite{TwoComponentMappingUkrain}.
\begin{figure}
    \centering
    \includegraphics[width=\linewidth]{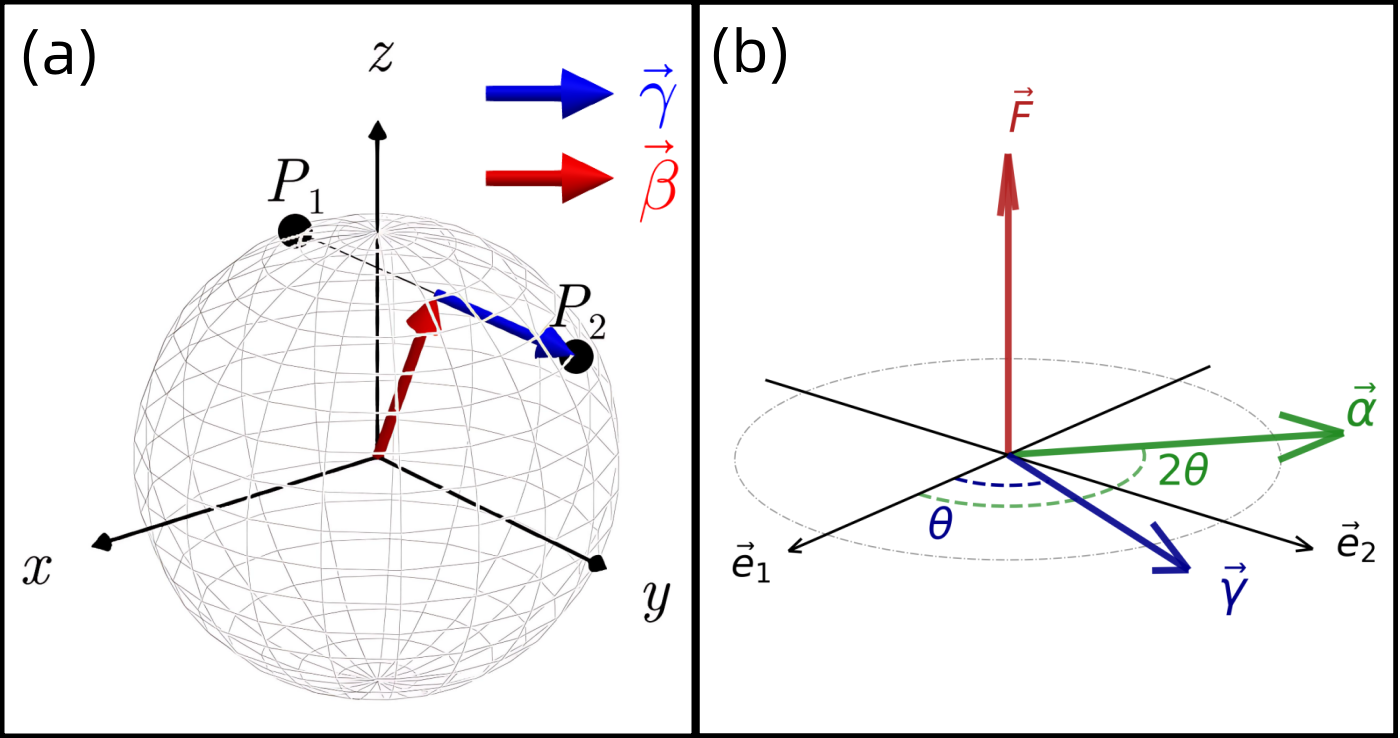}
    \caption{(a) Mapping a spin-1 state onto two Majorana stars, denoted as $\mathbf{P}_1$ and $\mathbf{P}_2$, on the Bloch sphere. The complete information of the spin-1 state is encoded in two key vectors: the average vector $\vec{\beta}$ (represented by the red arrow) and the relative vector $\vec{\gamma}$ (represented by the blue arrow). (b) The relative information can alternatively be expressed as a "vector" $\vec{\alpha}$. However, it's important to note that the $\vec{\alpha}$ vector transforms twice as fast as the actual relative vector $\vec{\gamma}$ under spin-rotation along $\vec{F}$. This $\vec{\alpha}$ vector effectively represents the expectation values of pseudo-spin operators with the second type subspace projection.}
    \label{Bloch_sphere}
\end{figure}
 
 \textit{Recast nematic order.--}The rotational properties of a relative vector $\vec{\gamma}$ are not directly accessible through standard spin-1 operators. However, it can be effectively captured using a state-dependent SU(2) reducible representation. This approach is particularly useful when analyzing systems with a preferred magnetization direction.

Consider a system with a magnetization direction defined by the unit vector $\vec{n}_{\beta} \equiv \vec{\beta}/|\vec{\beta}|$. Under rotations about this axis, the average vector $\vec{\beta}$ remains invariant. In contrast, the relative vector $\vec{\gamma}$ rotates within the plane $\Pi_\beta$ orthogonal to $\vec{n}_{\beta}$. This behavior implies that $\vec{\gamma}$ transforms according to a two-dimensional reducible representation of SO(2)$\subset$ SU(2) within the $\Pi_\beta$ plane.
To analyze this, we first define the direction of magnetization as $\mathbf{e}_3$. We then establish an orthonormal Cartesian coordinate frame $\mathbf{e}_1, \mathbf{e}_2, \mathbf{e}_3$. Let  $\sigma_i = \mathbf{e}_i \cdot \mathbf{\sigma}$ represent the Pauli matrices projected along the respective basis vectors, where $\mathbf{\sigma} = (\sigma_x,\sigma_y,\sigma_z)$ is the vector of Pauli matrices, and $I$ is the $2\times 2$ identity matrix. We define the following symmetric operators 
 \begin{eqnarray}
 \nonumber
 \hat{\alpha}_1 &\equiv   \frac{1}{2}(\sigma_1 \otimes \sigma_1 - \sigma_2 \otimes \sigma_2), \\
 \nonumber
 \hat{\alpha}_2 &\equiv \frac{1}{2}(\sigma_1 \otimes \sigma_2 + \sigma_2 \otimes \sigma_1), \\
 \hat{S}_3 &= \frac{1}{2}(\sigma_3 \otimes I + I \otimes \sigma_3).
 \end{eqnarray} 
 These three operators, $\hat{\alpha}_1,\hat{\alpha}_2,$ and $\hat{S}_3$ form a SU(2) reducible representation with structure constant 2. While $\hat{S}_3$ functions as a conventional spin operator, the expectation value of $\hat{\alpha}_1$ and $\hat{\alpha}_2$ exhibit a unique rotational behavior. Specifically, if a spin rotation of angle $\theta$ is applied along the $\mathbf{e}_3$ axis, the expectation values $\langle \hat{\alpha}_1\rangle$ and $\langle \hat{\alpha}_2 \rangle$ rotate by an angle of $2\theta$, See Fig.~\ref{Bloch_sphere}.(b). The invariant subspace of this representation is spanned by the eigenstate of $\hat{S}_3$ with an eigenvalue of 0. The expectation values of these operators can be expressed in terms of the components $\vec{\beta} =(\beta_1,\beta_2,\beta_3)$ and $\vec{\gamma}=(\gamma_1,\gamma_2,\gamma_3)$ as follows:
 \begin{eqnarray}
 \nonumber
     \langle \hat{\alpha}_1 \rangle &=& \frac{n_{\Phi}}{2}(\beta_1^2 - \beta_2^2) - \frac{n_{\Phi}}{2}(\gamma_1^2 - \gamma_2^2),\\
     \langle \hat{\alpha}_2 \rangle &=& n_{\Phi}(\beta_1 \beta_2 - \gamma_1 \gamma_2), \langle \hat{S}_3 \rangle = n_{\Phi}\beta_3.
     \label{Exp_red_operators}
 \end{eqnarray}
 In the current context, our focus is on the operators that act as the projections onto specific directions $\vec{\beta}=(0,0,\beta_3), \vec{\gamma} = (\gamma_1,\gamma_2,0)$. Within this constraint, the expectation values satisfy $\langle \hat{\alpha}_1\rangle^2 + \langle \hat{\alpha}_2\rangle ^2 + \langle \hat{S}_3\rangle^2 = n^2 $. We construct continuous transformation operators \(e^{i\theta \hat{\alpha}_1}\) and \(e^{i\theta \hat{\alpha}_2}\) respectively. The action of these operators differs from standard spin-rotation operators about the magnetization direction. Instead, they modify the angle between the two Majorana stars associated with the quantum state. The complete set of operators $\hat{S}_1, \hat{S}_2, \hat{S}_3, \hat{\alpha}_1, \hat{\alpha}_2$ provide tools to transform into arbitrary state within the spin-1 space. By strategically choosing $\mathbf{e}_3$ to align with $\mathbf{e}_x,\mathbf{e}_y,\mathbf{e}_z$, this construction generates all necessary nematic operators. When the spin-1 state is projected onto an effective two-component system, these operators, $\hat{\alpha_1},\hat{\alpha_2},\hat{S}_3$ reduce to the three Schwinger pseudo-spin operators, providing a direct mapping for analyzing the effective two-component dynamics within the spin-1 framework.

 \textit{Soliton dynamics within geometrical perspective.--}
The geometric perspective provides a unified lens to understand the existence and distinct manifestation of spin-1 topological solitons, which arise from $\mathbb{Z}_2$ degeneracies in the underlying geometric vector space of $\vec{\beta}$ and $\vec{\gamma}$.

Generally, the dynamics of an order parameter $\mathcal{O}$ (spin-order $\vec{\mathbf{F}}$ or nematic order $\vec{\Gamma}$) can be described by a continuity-like equation:
\begin{align}
    \frac{d \mathcal{O}}{dt} = -\frac{d}{dx} J^{O} + \text{source term}.
\end{align}
For a kink structure in $\mathcal{O}$, the first term of right hand side, representing the current $J^O$, primarily governs deformation, while the source term dictates the global movement. Conversely, for a dip in $\mathcal{O}$, the dynamics are predominantly driven by the current term, as propagation is not constrained by conservation laws.

 In ferromagnetic spin-1 BECs, a $\pi$- kink in the $\vec{\beta}$ vector corresponds to a Ferrodark soliton(FDS)\cite{XiaoquanYuFerrodark}. This is accompanied by a domain wall structure in the transverse magnetization. Crucially, the dynamics of FDSs are entirely determined by the quadratic Zeeman field, which acts as the source term in their equations of motion.

In antiferromagnetic condensates $g_s >0$, the quadratic Zeeman effect $q>0$ plays a crucial role in defining the ground state configuration.  The Zeeman energy, $E_{Z} = q n (1 + \beta_z^2 -\gamma_z^2)/2$ is minimized when the relative vector $\vec{\gamma}$ aligns with the magnetic field axis. This creates a ground state with a discrete $\mathbb{Z}_2$ degeneracy in vector space, corresponding to two energetically equivalent configurations $\vec{\gamma} = (0,0,+1)$ and $\gamma = (0,0,-1)$ while $\vec{\beta}$ remains zero.
\begin{figure}[h]
    \centering
    \includegraphics[width=\linewidth]{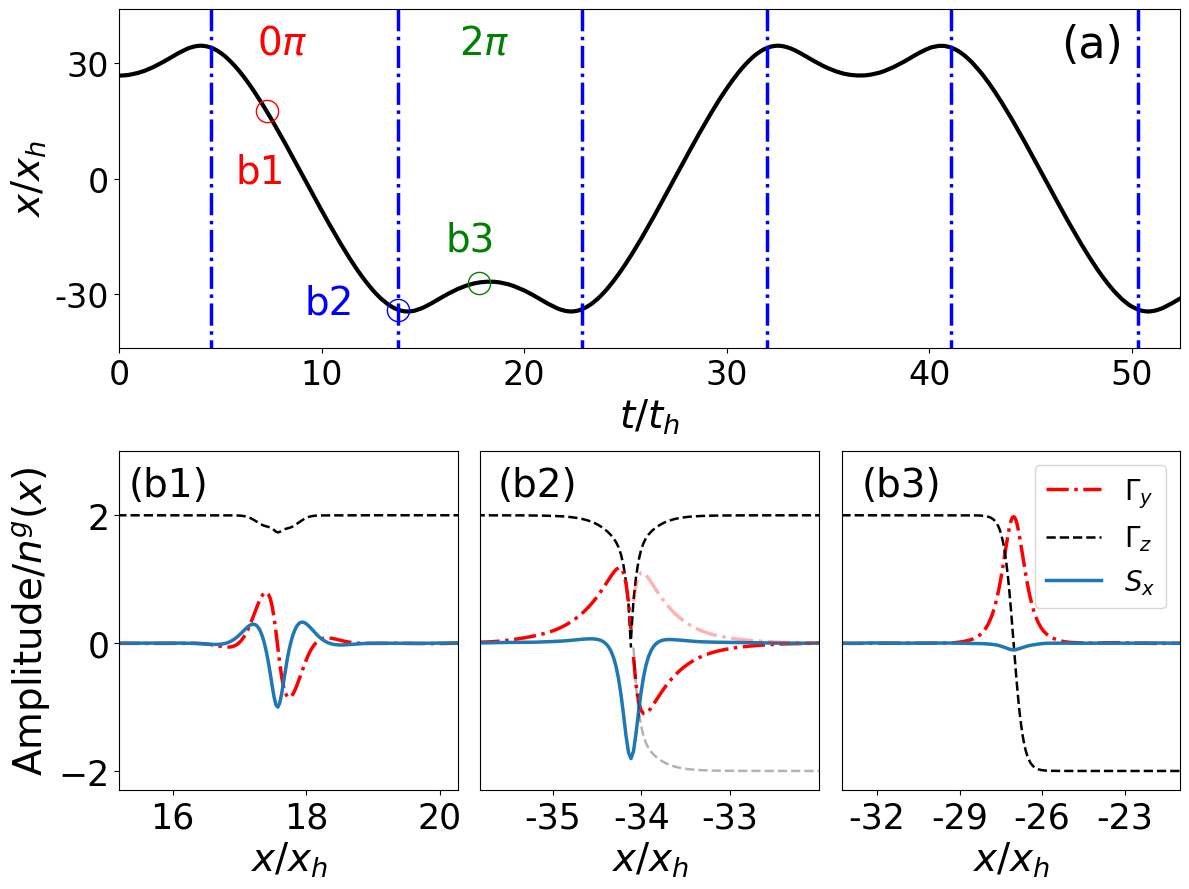}
    \caption{(a) The trajectory of a magnetic soliton within a harmonic trap with a quadratic Zeeman energy of $q=0.19g_s n_b^{\text{peak}}$, where $n_b^{\text{peak}}$ is the peak density of ground state. The trajectory and depicted transition were originally presented in Ref.~\cite{RelativePhaseDomainwall}. The blue dashed lines indicate the transition points where the magnetic soliton changes between $0\pi$ and $2\pi$ type. We have selected three points- b1,b2 and b3- to highlight the geometric structure corresponding to the 0$\pi$ state, the transition point, and the $2\pi$ state, respectively.Their geometric configurations are detailed in figures (b1), (b2), and (b3).  Since our simulations align the magnetic field along $x$-axis, we consider $\Gamma_y$, $\Gamma_z$and $S_x$ as the complete set of physical quantities. For the 0$\pi$ state shown in (b1), $\Gamma_z$ exhibits a small dip. In contrast, for the $2\pi$ state in (b3), $\Gamma_z$ forms a kink. At the transition point (b2), specifically at the soliton core, both $\Gamma_z$,$\Gamma_y$ become zero. This allows for a transformation where $\vec{\Gamma}$ on one side of the magnetic soliton effectively flips to $-\vec{\Gamma}$ on one sides of magnetic soliton (denoted by transparent lines), while maintaining the continuity of $\vec{\Gamma}$.This means that at the transition point, the dip and kink geometric configurations are equivalent under this transformation.}
    \label{magnetic_fig}
\end{figure}

A magnetic soliton can then form as a stable $\pi$-kink in the spatial profile of $\vec{\gamma}$ that connects these two degenerate vacua. Crucially, this kink structure is hidden from traditional order parameters. Because the total state is symmetric under the exchange of the two Majorana stars, a flip in the relative vector $\vec{\gamma} \rightarrow -\vec{\gamma}$ corresponds to a sign change in the singlet amplitude ($\phi_0\rightarrow -\phi_0$ ), leaving observables like density $n$, spin density $\mathbf{F}$ and nematic density $\mathcal{N}_{ij}$ unchanged. Consequently, the kink in $\vec{\gamma}$-space does not appear as a kink in any directly measurable density profile.
While the geometric method overcomes this limitation.  Typically, magnetic solitons are studied within a two-component theoretical framework. Here, we extend this understanding by embedding the magnetic soliton within the geometric framework of spin-1 condensates. In a one-dimensional configuration, the equations of motion for constructed vector $\Gamma_z$ in Eq.~\eqref{gamma_define}  provides a dynamical description of this structure\cite{footnote3}
\begin{eqnarray}
    \frac{\partial\Gamma_z}{\partial t} = -\frac{\partial}{\partial x}J_z^\Gamma +Q_g - g_s n_\Phi \kappa_z+ q(1+ \frac{|\psi_0|^2}{|\phi_0|^2})\kappa_z.
\label{EOM_GAMMA}
\end{eqnarray}
The explicit expressions of $J_z^\Gamma$ and $Q_g$ , along with their detailed derivations, are provided in Appendix. To clarify the dynamical implications of the geometric framework, we focus on the limit $q\ll g_s n^g$. The soliton width $\xi_s = \sqrt{\hbar/2mq}$ \cite{RelativePhaseDomainwall} set the scale of derivative coupling term $Q_g$, which remains small compared to the interaction-dominated source terms. Consequently, by integrating the source terms without spatial derivative, we can derive a relationship between the soliton velocity and the geometric structure. Within harmonic trap, magnetic soliton undergoes novel transition between $0\pi$ and $2\pi$ type\cite{RelativePhaseDomainwall}, see Fig.~\ref{magnetic_fig}. Their trajectory exhibit notable distinction. For $2\pi$ soliton, the velocity is directly given by the integral of source terms in Eq.~\eqref{EOM_GAMMA}. The geometric structure, $\Gamma_y$ and $F_x$ undergoes great change as the velocity change. In stark contrast, the $0\pi$ soliton, corresponding to a dip in $\Gamma_z$(see Fig.~\ref{magnetic_fig}(b1)). This non-topological structure is expected to behave not as a driven defect, but a quasi particle possessing constant effective mass as moving towards high-density regime. This is fundamentally different from the force driven motion of the kink.  From the geometric perspective, at the transition point, structures in $\vec{\Gamma}$ at the soliton core touch zero(equivalently, $|\mathbf{F}|=n$), the underlying wave function's symmetry allows for a crucial transformation: $\vec{\gamma}$ on one side of the zero point can effectively flip its sign, becoming $-\vec{\gamma}$. This maintains the continuity of the $\vec{\Gamma}$ spatial distribution. This flip transforms the $\Gamma_z$ profile from a dip structure into a kink structure, as it now effective connects two distinct "vacua" in $\vec{\gamma}$ space. This structural metamorphosis directly corresponds to the magnetic soliton undergoing a novel transition from a $0\pi$ to a $2\pi$ type. 

\textit{Conclusion and Discussion.--} In conclusion, we have established a formal correspondence between spin-1 geometry and a distinguishable two-particle Hilbert space. By embedding the physical spinor into this augmented manifold, we extract the relative geometric information of Majorana stars through the direct Hermitian operator. 

This geometric framework transforms our understanding of subspace projection as geometric constraints. Magnetic solitons reveal themselves as hidden kinks in $\vec{\gamma}$, and the observed $0\pi - 2\pi$ soliton transition corresponds to a fundamental geometric change between dip and kink configurations, reflecting the spin-1's underlying exchange symmetry. 
Our approach establishes a universal paradigm where quantum states are understood through elementary geometric vectors with distinct symmetry properties, providing both intuitive visualization and computational power for high-spin systems.

\textit{Acknowledgement.}-- We thank Y.X. Bai for the early-stage contribution and insightful discussions.


%

\pagebreak
\appendix
\singlespacing 

\widetext


\pagebreak
\widetext
\begin{center}
	\textbf{\large Supplemental Material for ``Unified Geometric perspective for Spin-1 Systems: Bridging Nematic Director and Majorana Stars''}
\end{center}

\setcounter{equation}{0}
\setcounter{figure}{0}
\setcounter{table}{0}
\makeatletter
\renewcommand{\theequation}{S\arabic{equation}}
\renewcommand{\thefigure}{S\arabic{figure}}


%

	This Supplemental Material includes the detailed proof of the equivalence between director and $\vec{\gamma}$, detailed derivation of subspace projection condition,  and projection of ferrodark soliton in spin-1 easy-plane phase, the Bogoliubov specturm relationship between spin-1 subject to quadratic Zeeman and two component condensate with coherent coupling,  the derivation of equation of $\vec{\Gamma}$ and other  necessary materials for supporting the main results presented in the main manuscript. 
	


	
	

\maketitle

\section{The equivalence between nematic director and relative vector $\vec{\gamma}$}

In this section, we provide a detailed derivation establishing the equivalence between the nematic director $\vec{u}$ and the relative vector $\vec{\gamma}$ of two Majorana stars in our formulation. This equivalence is central to interpreting the physical meaning of $\vec{\gamma}$ as the primary orientational axis of the system.
We begin by considering the complex vector order parameter $\vec{\psi}$ in Cartesian representation, and decompose it into real and imaginary parts:
\bea
\vec{\psi} = \vec{a} + i \vec{b}
\eea
where $\vec{a}$,$\vec{b} \in \mathbb{R}^3$. Since the overall U(1) phase of $\vec{\psi}$ is physically irrelevant for nematic director, we can perform a phase rotation to bring $\vec{\psi}$ into a canonical form that isolates the nematic director explicitly. Specifically, we write
\bea
\vec{\psi} = e^{i\theta}(\vec{u}+i\vec{v}),
\eea
where $\theta \in [0,2\pi)$ is a global phase chosen such that the resulting real vectors $\vec{u}$ and $\vec{v}$ satisfy two key conditions:
\bea
\vec{u}\cdot \vec{v} = 0, \text{and} |\vec{u}| > |\vec{v}|.
\eea
Expanding the phase-rotated expression yields the relations
\bea
\vec{u} = \cos \theta \vec{a}+ \sin\theta \vec{b}, \vec{v} = -\sin\theta \vec{a} + \cos \theta \vec{b}.
\eea
Carrying out the dot product, we obtain
\bea
\tan 2\theta = \frac{2\vec{a}\cdot \vec{b}}{\vec{a}\cdot\vec{a}-\vec{b}\cdot\vec{b}}.\label{theta_relation1}
\eea
Now, within distinguishable spin-1/2 product-state framework, we have shown in the main text that
\bea
\vec{\psi} = \frac{1}{\phi_0^*}(\vec{\Gamma}+ i \vec{\kappa}),
\eea
crucially, the singlet amplitude satisfies:
\bea
\phi_0^2 = \vec{\psi} \cdot \vec{\psi} = |\vec{a}|^2 - |\vec{b}|^2 + 2i \vec{a}\cdot \vec{b}. \label{phi0_2}
\eea
Let us write $\phi_0 = |\phi_0|e^{i\theta_0}$, comparing Eq.~\eqref{phi0_2} with the form of a complex number, we find
\bea
\tan (2\theta_0) = \frac{2\vec{a}\cdot \vec{b}}{|\vec{a}|^2- |\vec{b}|^2}.
\eea
This identical to Eq.\eqref{theta_relation1}, which $\theta = \theta_0 + \pi/2$, and 
\bea
\vec{u} = \frac{\vec{\Gamma}}{|\phi_0|}, \vec{v} = \frac{\vec{\kappa}}{|\phi_0|}.
\eea
\vspace{5cm}

\section{The subspace projection and geometric constraint}
In the main context, we mentioned there are two types of subspace space projection. In this section, we show how they connect to geometric constraint. 
Expressing component wave functions in terms of polar angles$\theta_i$ and  azimuthal angles $\phi_i$, with $i=1,2$. To focus on the internal geometric structure, we introduce a normalization constant $C$ and omit the global phase:

\bea
\psi_{+1} &=& \sqrt{2} C \cos \frac{\theta_1}{2} \cos \frac{\theta_2}{2},\\
\psi_0 &=& C(\cos \frac{\theta_1}{2} \sin \frac{\theta_2}{2} e^{i\phi_2} + \cos \frac{\theta_2}{2} \sin \frac{\theta_1}{2} e^{i\phi_1}),\\
\psi_{-1} &=& \sqrt{2}C \sin \frac{\theta_1}{2} \sin \frac{\theta_2}{2} e^{i(\phi_1 + \phi_2)}.
\eea
Here $C$ ensures the total density $\sum_{i=0,\pm 1} |\psi_i|^2 = n$ is conserved. The first type of subspace projection, $\psi_{+1}$($\psi_{-1}$) $=0$ corresponds to at least one star located at south(north) pole,i.e.  $\theta_1$ or $\theta_2 = \pi$($0$). 
The second type of subspace projection, $\psi_0=0$, requires both the real part and the imaginary part equal to zero. Expressing $\psi_0 = 0$ yields the conditions:
\bea
\cos \frac{\theta_1}{2} \sin \frac{\theta_2}{2}e^{i\phi_2} = - \cos \frac{\theta_2}{2} \sin \frac{\theta_1}{2} e^{i\phi_1}.
\eea
The only way to satisfy both equations (within physical ranges restrict $\theta_i \in [0,\pi]$ ) is to impose a phase locking condition:
\bea
\phi_1 = \phi_2 + \pi \quad (\text{mod } 2\pi)
\eea
together with an amplitude balance:
\bea
\cos\frac{\theta_1}{2} \sin \frac{\theta_2}{2} = \cos \frac{\theta_2}{2} \sin \frac{\theta_1}{2}.
\eea
The latter simplifies to $\tan \frac{\theta_1}{2} = \tan \frac{\theta_2}{2}$, implying $\theta_1 = \theta_2$. Thus, the condition $\psi_0 = 0$ enforces a geometric configuration on the two Marjorana stars: they must have equal polar angles and opposite azimuthal phases, thus antipodally symmetric about $z-$axis on the Bloch sphere.

\section{Subspace projection of soliton solutions}
As an illustrative example, we demonstrate how the ferro-dark soliton(FDS) in easy-plane phase can be mapped into a two component system with coherent coupling. Other spin-1 excitations satisfying $\psi_{+1} = \psi_{-1}$ are similarly equivalent to FDS solution. A general mapping requires verify the condition in main manuscript.
 
The spinor $\psi = (\psi_{+1},\psi_0,\psi_{-1})^T$ evolves according to the Gross-Pitaevskii equation:
\begin{align}
     i\hbar \frac{\partial \psi_{\pm 1}}{\partial t} &= H_0 \psi_{\pm 1} + g_s (n_0 + n_{\pm 1} - n_{\mp 1})\psi_{\pm 1} + q\psi_{\pm 1}+ g_s \psi_0^2 \psi_{\mp 1}^*,
     \label{spin1_GPE1}
     \\
     i\hbar \frac{\partial \psi_0}{\partial t} &= H_0 \psi_0 + g_s(n_{+1} + n_{-1})\psi_0 + 2g_s \psi_0^* \psi_{+1} \psi_{-1},
     \label{spin1_GPE2}
 \end{align} 
where $H_0 \equiv -\frac{\hbar^2 \nabla^2}{2M} + g_n n$, $n_i = |\psi_i|^2, n = \sum_{i=0,\pm 1} n_i$ and $q$ is the quadratic Zeeman energy, $g_n$ is the density dependent interaction strength and $g_s$ is the spin-dependent interaction strength. In one dimension, for the parameters $g_n=-2g_s>0$, $0<q<-2g_s n_b$, this equation admit a class of exact solutions known as ferro-dark solitons\cite{XiaoquanYuFerrodark}. FDSs are classified into type-$\rm I$ and type-$\rm II$ based on  the sign of their inertial mass. We focus on the type-$\rm I$ branch to illustrate the subspace projection procedure.  The wavefunction of a propagating type-$\rm I$ FDS is given by \cite{bai2025FDS_collision}
\begin{align}
    \psi_{\pm 1}^{\rm I}(x,t) &= \sqrt{n_{\pm 1}^g}[-\alpha \tanh(\xi/\ell) + i \delta],\\
    \psi_0^{\rm I}(x,t) &= \sqrt{n_0^g}[\alpha - i \delta \tanh(\xi/\ell)].
\end{align}
Here $\xi = x-Vt$, $V$ is the soliton's propagating velocity, and $n_i^g$ represents the  background density of $i-$th component, with $n_{\pm 1}^g = n^g (1-\tilde{q})/4 and n_0^g = n^g(1+\tilde{q})/2$. The parameter $\tilde{q} = q/(2|g_s|n^g)$, and $n^g = \sum_i n_i^g$. The soliton width $\ell = \sqrt{2\hbar^2/(g_n n^g - MV^2-Q)}$, where $Q = \sqrt{M^2 V^4 +q^2-2g_n n^g MV^2}$. The coefficients $\alpha$ and $\delta$ are defined as $\alpha = \sqrt{(q+MV^2+Q)/2q}$ and $\delta = \sqrt{(q-MV^2-Q)/2q}$. This FDS solution exhibits a finite magnetization aligns along $x$-axis, with $F_x = \sqrt{(n^g)^2 - q^2/g_n^2} \tanh \left(\xi / \ell\right)$and $F_y = F_z = 0$. Notably, the wavefunction satisfies $\psi_{+1} = \psi_{-1}$. To map this state onto an effective two-component system, we apply a spin-rotation matrix $U \equiv e^{i\pi/2 S_y}$, explicitly given by
\begin{align}
    U = \begin{pmatrix}
        \frac{1}{2} & \frac{1}{\sqrt{2}} & \frac{1}{2}\\
        -\frac{1}{\sqrt{2}} & 0 & \frac{1}{\sqrt{2}}\\
        \frac{1}{2} & -\frac{1}{\sqrt{2}} & \frac{1}{2}
    \end{pmatrix}.
\end{align}
Under this rotation $\tilde{\psi} = U \psi$, the $m_F =0$ component wavefunction vanishes, i.e., $\tilde{\psi}_0 = (\psi_{-1}-\psi_{+1})/\sqrt{2} =  0$. The Hamiltonian transforms according to $H' = UHU^{-1}$. By imposing the condition $\tilde{\psi}_0 = 0$, the equations for $\tilde{\psi}_{+1}$ and $\tilde{\psi}_{-1}$ reduce to
\begin{align}
\label{Two_component_1}
    i\hbar \frac{\partial \tilde{\psi}_{+1}}{\partial t} &= -\frac{\hbar^2}{2m}\frac{d^2}{dx^2} \tilde{\psi}_{+1} + (g_n+g_s) \tilde{n}_{+1} \tilde{\psi}_{+1} + (g_n-g_s)  \tilde{n}_{- 1} \tilde{\psi}_{-1} +  \frac{q}{2}(\tilde{\psi}_{+ 1}+ \tilde{\psi}_{-1}),\\
    \label{Two_component_2}
    i\hbar \frac{\partial \tilde{\psi}_{-1}}{\partial t} &=  -\frac{\hbar^2}{2m}\frac{d^2}{dx^2} \tilde{\psi}_{-1} + (g_n+g_s)  \tilde{n}_{-1}\tilde{\psi}_{-1} + (g_n -g_s) \tilde{n}_{+ 1} \tilde{\psi}_{-1} + \frac{q}{2}(\tilde{\psi}_{+ 1}+ \tilde{\psi}_{-1}).
\end{align}
which describe a two-component condensate with coherent coupling\cite{Stringari_two_component}. The rotated wavefunction
\bea
\tilde{\psi}_{+1} = \sqrt{n_{+1}^g}[-\alpha \tanh (\xi/\ell)+i\delta] + \sqrt{n_0^g/2}[\alpha - i\delta \tanh(\xi/\ell)],\\
\tilde{\psi}_{-1} = \sqrt{n_{+1}^g}[-\alpha \tanh (\xi/\ell)+i\delta] - \sqrt{n_0^g/2}[\alpha - i\delta \tanh(\xi/\ell)],
\eea
constitute a class of exact solution to Eqs.~\eqref{Two_component_1} and \eqref{Two_component_2}.
\section{Bogoliubov-De-Genns spectrum}

In the study of low-energy excitations via the Bogoliubov-de Gennes(BdG) formalism, the structure of the underlying ground state plays a fundamental role. For a spin-1 Bose-Einstein condensate subject to a quadratic Zeeman field $q$, the BdG equations depend explicitly on the mean-field ground state $\tilde{\psi} ^g=\left(\tilde{\psi}_{+1}^g,\tilde{\psi}_0^g,\tilde{\psi}_{-1}^g \right)^T$. As noted in the main manuscript, any spatially uniform spin-1 state can be transformed into a representation where the $m=0$ component vanishes through a global spin rotation. We consider the easy-plane phase, which occurs in the parameter regime defined by $g_n>0,g_s<0,g_n > -g_s$ and $q<-2g_s n_b$\cite{Spin1GroundState}, where $n^g$ is the total density. In the laboratory frame (with the z-axis as the quantization aixs), the ground state component densities are $n_{\pm 1}^g = n^g (1-\tilde{q})/4, n_0^g = n^g (1+\tilde{q})/2$ where $\tilde{q}=q/(2|g_s| n^g)$. In this phase, the transverse magnetization is $|\mathbf{F}_\perp|=n^g \sqrt{1-\tilde{q}^2}$, and $F_z = 0$. The geometric order parameters given by Eq.~\eqref{gamma_define} representing relative information  are $\Gamma_z = \pm n^g \tilde{q}$ and $\Gamma_{x}=\Gamma_y=0$.

To simplify the excitation analysis, we perform a global spin rotation that aligns the transverse magnetization with the new quantization axis. In this rotated frame--which we adopt for the remainder of this derivation--the ground state simplifies to
\bea
\psi^g = \left(\sqrt{n_{+1}^g},0,\sqrt{n_{-1}^g}\right)^T,
\eea
where the rotated densities are $n_{+1}^g = n^g(1+\sqrt{1-\tilde{q}^2})/2$ and $n_{-1}^g = n^g(1-\sqrt{1-\tilde{q}^2})/2$.

To derive the excitation spectrum, we introduce small fluctuations around the ground state $\psi_i(\mathbf{r},t) = \psi_i^g + \delta \psi_{i,\mathbf{k}}(\mathbf{r},t)$, where the fluctuations take the form $\delta \psi_{i,\mathbf{k}}(\mathbf{r},t) = u_ie^{i(\mathbf{k}\cdot \mathbf{r}-\omega t)}+ v_i^* e^{-i(\mathbf{k}\cdot\mathbf{r}-\omega t)}$. This leads to the following eigenvalue problem
\bea
\hbar \omega \begin{pmatrix}
\mathbf{u}\\
\mathbf{v}
\end{pmatrix} = \mathcal{ L} \begin{pmatrix}
\mathbf{u}\\
\mathbf{v}
\end{pmatrix},\mathcal{L} = \begin{pmatrix}
    L+Q & M\\
    -M^* & -(L+Q)^*
\end{pmatrix}.
\eea
with the BdG matrix structured as
\bea
\mathcal{L} = \begin{pmatrix}
    L+Q & M\\
    -M^* & -(L+Q)^*
\end{pmatrix}.
\eea
Here, the quadratic Zeeman contribution is
\bea
Q = \begin{pmatrix}
    \frac{q}{2} & 0 & \frac{q}{2}\\
    0 & q & 0\\
    \frac{q}{2} & 0 & \frac{q}{2}
\end{pmatrix}
\eea
while the interaction and kinetic contribution are contained in $L$ and $M$, whose explict forms are
\bea
\nonumber
L = \begin{pmatrix}
    \epsilon_k + (g_n+g_s) (n^g+n_{+1}^g) - 2g_s n_{-1}^g & (g_n+g_s) \psi_0^g \psi_{+1}^g + 2g_s \psi_0^g \psi_{-1}^g & (g_n-g_s)\psi_{+1}^g \psi_{-1}^g\\
    (g_n+g_s) \psi_0^g \psi_{+1}^g + 2g_s \psi_0^g \psi_{-1}^g & \epsilon_k + (g_n+g_s) (n^g + n_0^g) - 2g_s n_0^g & (g_n + g_s) \psi_0^g \psi_{-1}^g + 2 g_s \psi_0^g \psi_{+1}^g\\
    (g_n - g_s)\psi_{-1}^g \psi_{+1}^g &  (g_n+g_s)\psi_0^g \psi_{-1}^g + 2g_s \psi_0^g \psi_{+1}^g & \epsilon_k + (g_n +g_s)(n^g + n_{-1}^g) - 2g_s n_{+1}^g
\end{pmatrix}
\eea

\bea
M = \begin{pmatrix}
    (g_n + g_s) n_{+1}^g & (g_n + g_s)\psi_0^g \psi_{+1}^g & (g_n - g_s)\psi_{+1}^g \psi_{-1}^g + g_s n_0^g\\
    (g_n+g_s) \psi_0^g \psi_{+1}^g & g_n n_0^g + 2g_s \psi_{+1}^g\psi_{-1}^g & (g_n+g_s)\psi_0^g \psi_{-1}^g\\
    (g_n-g_s) \psi_{+1}^g \psi_{-1}^g + g_s n_0^g & (g_n +g_s) \psi_0^g \psi_{-1}^g & (g_n+g_s) n_{-1}^g
 \end{pmatrix}
\eea
Now imposing the constraint $\psi_0^g = 0$  simplifies these matrices. All off-diagonal elements coupling the $m=0$ component to $m=\pm 1$ vanish, and the BdG matrix becomes block-diagonal upon reordering the basis as $(u_{+1},v_{+1},u_{-1},v_{-1},u_0,v_0)$
\bea
\mathcal{L} = \begin{pmatrix}
    \mathcal{L}^{(2)} & 0 \\
    0 & \mathcal{L}^{(0)}
\end{pmatrix}
\eea

The $4\times 4$ block $\mathcal{L}^{(2)}$ describes fluctuations within the $\{ |m=+1\rangle , |m=-1\rangle\}$ subspace and is identical to the BdG matrix of a two-component BEC given in Eqs.~\eqref{Two_component_1} and \eqref{Two_component_2}. The remaining $2\times 2$ block governs excitations that populate the $m=0$ component,
\bea
\mathcal{L}^{(0)} =\begin{pmatrix}
 \epsilon_k + \frac{q}{2} & -2g_s \sqrt{n_{+1}^g n_{-1}^g} \\
 2 g_s\sqrt{n_{+1}^g n_{-1}^g} & -\epsilon_k -\frac{q}{2}
\end{pmatrix}
\eea
This mode corresponds to a fluctuation that breaks the co-alignment condition of $\mathbf{F}$ ,thereby rotating the direction of magnetization out of the co-alignment direction.  This mapping provides interpretation of the spin-1 BdG spectrum of ferromagnetic phase in \cite{BdGspin1} and  easy-plane phase in supplement material of \cite{XiaoquanYuFerrodark} exhibit similiar behavior with the two component BdG spectrum \cite{Polar_BdG_twocomponent} with substitution $g_0 = g_n+g_s, g_1 = g_n - g_s$, $\Omega = \frac{q}{2}$ where $g_0$ is the intra-species interaction strength and $g_1$ is the inter-species interaction strength. $\Omega$ is the coherent coupling between the two components.

The similiar relationship also happens in polar phase in spin-1 BdG sprectrum \cite{BdGspin1} and two-component at neutral phase\cite{TwoComponentBdG}.

\section{The derivation of Equations of motion of $\vec{\Gamma}$}
In the main manuscript, we asserted that our formalism  directly yields the equations of motion for the vector  $\vec{\Gamma}$. Here, we present a detailed derivation for its $z$-component, $\Gamma_z = \phi_0^* \psi_0 + \phi_0 \psi_0^*$; the equations for the transverse components $\Gamma_x$ and $\Gamma_y$ follow an analogous procedure.
 
We begin with the mean-field dynamics of the spin-1 BEC, governed by the coupled GPEs provided by Eqs.~\eqref{spin1_GPE1} and \eqref{spin1_GPE2}. The time evolution of $\Gamma_z$ is given by the product rule:
\bea
i\hbar \frac{\partial \Gamma_z}{\partial t} = \left(i\hbar \frac{\partial \phi_0^*}{\partial t}\right) \psi_0 + \phi_0^* \left(i\hbar \frac{\partial \psi_0}{\partial t}\right) + \left(i\hbar \frac{\partial \phi_0}{\partial t}\right)\psi_0^* + \phi_0 \left(i\hbar \frac{\partial \psi_0^*}{\partial t}\right),\label{dGamma_t}
\eea
The dynamics of the condensate components are driven by the kinetic energy operator $\hat{K}=-\frac{\hbar^2}{2m}\nabla^2$ and interaction terms. Specifically, the equation for $\psi_0$ is:
\bea
i\hbar \frac{\partial \psi_0}{\partial t} = \hat{K}\psi_0 + g_n n \psi_0 + g_s(n_{+1}+n_{-1}-n_0)\psi_0 + 2g_s \psi_{+1}\psi_{-1}\psi_0^*.\label{dpsi0_dt}
\eea
To evaluate Eq.~\eqref{dGamma_t}, we first derive the equation of motion for the singlet amplitude $\phi_0$. Recall the definition $\phi_0^2 = \psi_0^2 - 2\psi_{+1}\psi_{-1}$. Differentiating this relation with respect to time yields:
\bea
\phi_0 \left(i\hbar \frac{\partial \phi_0}{\partial t}\right) = \psi_0 \left(i\hbar \frac{\partial \psi_0}{\partial t}\right)-\psi_{+1}\left(i\hbar \frac{\partial \psi_{-1}}{\partial t}\right) - \psi_{-1}\left(i\hbar \frac{\partial \psi_{+1}}{\partial t}\right).
\eea
Substituting the GPEs for $\psi_{0,\pm 1}$ into the right-hand side, we observe a cancellation of the spin-dependent interaction terms. Specifically, the terms proportional to $g_s$ sum to zero:
\bea
\left[\psi_0 \left(i\hbar \frac{\partial \psi_0}{\partial t}\right)\right]_{g_s} - \left[\psi_{+1} \left(i\hbar \frac{\partial \psi_{-1}}{\partial t}\right)\right]_{g_s} - \left[\psi_{-1}\left(i\hbar \frac{\partial \psi_{+1}}{\partial t}\right)\right]_{g_s}\\
=g_s\left[(n_{+1}+n_{-1})\psi_0^2 + 2 \psi_{+1}\psi_{-1} |\psi_0^2| - \left(n_0 + n_{-1}-n_{+1} + n_0 + n_{+1}-n_{-1}\right)\psi_{+1}\psi_{-1}\right]=0
\eea
The remaining contributions arise from the kinetic energy, the density-dependent interaction, and the quadratic Zeeman shift. The equation for $\phi_0$ simplifies to
\bea
i\hbar \frac{\partial \phi_0}{\partial t} = \frac{1}{\phi_0} \mathcal{K}[\psi] + g_n n \phi_0 - q \frac{2\psi_{+1}\psi_{-1}}{\phi_0},\label{d_phi0_dt}
\eea
where $\mathcal{K}[\psi] \equiv \psi_0 \hat{K}\psi_0 - \psi_{+1}\hat{K}\psi_{-1} - \psi_{-1}\hat{K}\psi_{+1}$ represents the kinetic contribution. Using the identity $2\psi_{+1}\psi_{-1} = \psi_0^2 -\phi_0^2$, the last term in Eq.~\eqref{d_phi0_dt} becomes $-q(\psi_0^2/\phi_0 - \phi_0)$.

We now substitute Eqs.~\eqref{d_phi0_dt} and \eqref{dpsi0_dt} into Eq~.\eqref{dGamma_t}. The density-dependent interaction terms proportional to $g_n$ cancel explicitly:
\bea
-\psi_0 g_n n \phi_0^* + \phi_0^* g_n n \psi_0 + \psi_0^* g_n  n \phi_0 - \phi_0 g_n n \psi_0^*=0.
\eea
The spin-dependent terms $\mathcal{T}_{\text{int}}$ proportional to $g_s$ survive
\begin{align}
\mathcal{T}_{\text{int}}&=\phi_0^* g_s (n_{+1} + n_{-1})\psi_0 + 2 \phi_0^* g_s \psi_0^* \psi_{+1}\psi_{-1}-\phi_0 g_s (n_{+1}+n_{-1})\psi_0^* - 2\phi_0 g_s \psi_0 \psi_{+1}^* \psi_{-1}^*\\
&= g_s (n_{+1}+n_{-1})\phi_0^* \psi_0  + g_s \phi_0^* \psi_0^* (\psi_0^2 -\phi_0^2) - g_s (n_{+1}+ n_{-1})\phi_0 \psi_0^*- g_s \phi_0 \psi_0\left((\psi_0^*)^2 - (\phi_0^*)^2\right)\\
&=g_s n \phi_0^* \psi_0 - g_s |\phi_0|^2 \phi_0 \psi_0^* - g_s n \phi_0 \psi_0^* + g_s |\phi_0|^2\psi_0 \phi_0^*\\
&=g_s (n+ |\phi_0|^2) (\phi_0^* \psi_0 - \phi_0 \psi_0^*).
\end{align}
The quadratic Zeeman term $\mathcal{T}_q$ combines as
\begin{align}
    \mathcal{T}_q &= \psi_0 q\frac{2\psi_{+1}^*\psi_{-1}^*}{\phi_0^*} - \psi_0^* q \frac{2\psi_{+1}\psi_{-1}}{\phi_0}\\
    &=q \psi_0 \frac{\left((\psi_0^*)^2 - (\phi_0^*)^2\right)}{\phi_0^*}- q\psi_0^* \frac{\left(\psi_0^2 -\phi_0^2\right)}{\phi_0}\\
    &=q \frac{n_0}{|\phi_0|^2}\psi_0^*\phi_0-q \psi_0 \phi_0^* -q \frac{n_0}{|\phi_0|^2} \psi_0 \phi_0^* + q \psi_0^* \phi_0\\
    &=q (1+ \frac{n_0}{|\phi_0|^2}) (\psi_0^* \phi_0 - \psi_0 \phi_0^*).
\end{align}
Finally, we collect the kinetic energy term $\mathcal{T}_{\text{kin}}$. The combination of spatial derivatives take the form of the divergence of a current density $J_z^\Gamma$ and a geometric source term $Q_g$
\begin{align}
\mathcal{T}_{\text{kin}} = -\frac{\psi_0}{\phi_0^*}(\mathcal{K}[\psi])^* + \phi_0^* \hat{K} \psi_0 + \frac{\psi_0^*}{\phi_0} \mathcal{K}[\psi] - \phi_0 \hat{K}\psi_0^*\\
= -\frac{\psi_0}{\phi_0^*}(\psi_0^* \hat{K}\psi_0^* - \psi_{+1}^* \hat{K}\psi_{-1}^* - \psi_{-1}^* \hat{K}\psi_{+1}^*) + \phi_0^* \hat{K}\psi_0 + \frac{\psi_0^*}{\phi_0}(\psi_0 \hat{K} \psi_0 - \psi_{+1}\hat{K}\psi_{-1} - \psi_{-1}\hat{K}\psi_{+1}) - \phi_0 \hat{K} \psi_0^*
\end{align}
Making use of the identity $A\nabla^2 B = \nabla\cdot (A \nabla B) - (\nabla A) \cdot(\nabla B)$, and the relation $\psi_0 \nabla \psi_0 - \psi_{+1}\nabla \psi_{-1} - \psi_{-1}\nabla \psi_{+1} = \phi_0\nabla \phi_0$, we analyze the term
\begin{align}
   & \frac{\psi_0^*}{\phi_0}(\psi_0 \hat{K} \psi_0 - \psi_{+1}\hat{K}\psi_{-1} - \psi_{-1}\hat{K}\psi_{+1}) \\
    &= -\frac{\hbar^2}{2m} \nabla \cdot\left[\frac{\psi_0^*}{\phi_0} \left(\psi_0 \nabla \psi_0 - \psi_{+1}\nabla \psi_{-1} - \psi_{-1}\nabla \psi_{+1}\right)\right] + \frac{\hbar^2}{2m}\left[\left(\nabla\frac{n_0}{\phi_0}\right)\cdot \left(\nabla\psi_0\right)- \left(\nabla \frac{\psi_0^* \psi_{+1}}{\phi_0}\right)\cdot(\nabla \psi_{-1})- \left(\nabla \frac{\psi_0^* \psi_{-1}}{\phi_0}\right)\cdot \left(\nabla \psi_{+1}\right)\right]\\
    &=-\frac{\hbar^2}{2m} \nabla \cdot (\psi_0^* \nabla \phi_0) + \frac{\hbar^2}{2m}\left[\left(\nabla\frac{n_0}{\phi_0}\right) \cdot\left(\nabla\psi_0\right)- \left(\nabla \frac{\psi_0^* \psi_{+1}}{\phi_0}\right)\cdot(\nabla \psi_{-1})- \left(\nabla \frac{\psi_0^* \psi_{-1}}{\phi_0}\right)\cdot \left(\nabla \psi_{+1}\right)\right],\\
    &= \psi_0^* \hat{K}\phi_0 + \frac{\hbar^2}{2m}\left[\left(\nabla\frac{n_0}{\phi_0}\right) \cdot\left(\nabla\psi_0\right)- \left(\nabla \frac{\psi_0^* \psi_{+1}}{\phi_0}\right)\cdot(\nabla \psi_{-1})- \left(\nabla \frac{\psi_0^* \psi_{-1}}{\phi_0}\right)\cdot \left(\nabla \psi_{+1}\right) - (\nabla\psi_0^*) \cdot (\nabla \phi_0)\right].
\end{align}
We denote the source term arising from the gradient as
\begin{align}
    Q_g = \frac{\hbar}{2im} \left[\left(\nabla\frac{n_0}{\phi_0}\right)\left(\nabla\psi_0\right)- \left(\nabla \frac{\psi_0^* \psi_{+1}}{\phi_0}\right)(\nabla \psi_{-1})- \left(\nabla \frac{\psi_0^* \psi_{-1}}{\phi_0}\right) \left(\nabla \psi_{+1}\right) - (\nabla \psi_0^*)\cdot (\nabla \phi_0)\right]-c.c.
\end{align}
the current density
\begin{align}
J_z^\Gamma = -\frac{\hbar}{2im}  (\psi_0^* \nabla \phi_0 - \phi_0 \nabla \psi_0^* + \phi_0^* \nabla \psi_0 - \psi_0 \nabla \phi_0^*).
\end{align}
Combining the results for $\mathcal{T}_{\text{int}}$, $\mathcal{T}_q$ and $\mathcal{T}_{\textbf{kin}}$, we arrive at the complete equation of motion for $\Gamma_z$ as presented in Eq.~\eqref{EOM_GAMMA} of the main text.


%




\end{document}